\documentclass[pre,aps,amsmath,twocolumn,amssymb,showpacs]{revtex4}%

\usepackage{amsmath}
\usepackage{amsfonts}
\usepackage{amssymb}
\usepackage{graphicx}
\usepackage{dcolumn}
\usepackage{bm}
\usepackage{hyperref}
\usepackage{color}
\begin{document}
\preprint{}
\title{On the relevance of numerical simulations to booming sand}

\author{Patrick Richard, Sean McNamara and Merline Tankeo}
\affiliation{Institut de Physique de Rennes, Universit\'e de Rennes I, UMR CNRS 6251, F-35042 Rennes, France}
\date{\today}
\begin{abstract}
We have performed a 
simulation study of 3D cohesionless granular flows  down an inclined chute. We find that the 
oscillations observed 
in [L.~E. Silbert, {\em Phys. Rev. Lett.}, {\bf{94}} 098002 (2005)]
near the angle
of repose are harmonic vibrations
of the lowest normal mode. 
Their frequencies depend on the contact stiffness 
as well as 
on the depth of the flow. 
Could these oscillations account for the phenomena of ``booming sand''?
We estimate an effective contact stiffness from the Hertz law,
but this leads to frequencies several times higher than observed.
However, the Hertz law also predicts interpenetrations of a few nanometers, 
indicating that the oscillations frequencies are 
 governed by the surface stiffness, which can be much lower than the bulk one. 
This is in agreement with previous studies ascribing the ability to sing to the  
presence of a soft coating on the grain surface.
\end{abstract}
\maketitle
A ``booming'' or ``singing'' dune 
is a sand dune that emits a loud sound
when an avalanche occurs on its slip face.
The sound can be very loud -- audible up to 10 km away --
and has a well defined frequency
of order one hundred Hertz.
The physical origins of this phenomenon are still matter of debate
in spite of much experimental and theoretical work
\cite{Andreotti2008,Andreotti2009,Douady2006,Vriend2007,Andreotti2004,Mills2009}.
A successful theory of booming sand must explain 
why the sound is not a mix of a wide range of frequencies,
and therefore several frequency selection mechanisms have been proposed.
We now summarize them non-exhaustively.
According to Andreotti~\cite{Andreotti2004}, for a given grain size, the frequency is 
set by the shear rate of the shear band separating the avalanche to the static part of the dune. 
This first explanation is not compatible with the one proposed by Douady et al.~\cite{Douady2006}, in which the frequency is set by a resonance of the flowing layer.
A third explanation has been proposed by Vriend et al.~\cite{Vriend2007}: 
The frequency is not selected by the properties of the avalanche but by the acoustical resonance induced by the stratification of the dune, explaining why the frequency may vary with time.  More recently~\cite{Andreotti2009}, Andreotti and Bonneau show that the shear band separating an avalanche from the static part of the dune induces an amplification of guided elastic waves, leading to a linear instability. The frequency is then set by the maximum value of the instability growth rate. 
Finally Mills and Chevoir~\cite{Mills2009} 
made the interesting remark
that something like booming sand had already been observed 
in numerical simulations \cite{Silbert_PRL_2005} that exhibited spontaneous oscillations
in granular flows near the angle of repose.
These oscillations were interpreted as signs of intermittency
near the jamming transition.

Here,
we examine 
Mills and Chevoir's re-interpretation of these
oscillations as an acoustical phenomena -- possibly ``booming sand''.
We extend the simulations performed in~\cite{Silbert_PRL_2005} 
by carrying out an extensive study 
of the influence of the contact normal stiffness 
and of the height of the flow.
After a description of the method,
we will show that the observed oscillations
are multi-body harmonic oscillations.
Their frequency is thus governed
by the stiffness of the springs 
that model the repulsive inter-granular forces.
We estimate a reasonable value for this quantity
and discuss the relevance of the simulations.

\textit{Method.}
We use our own implementation~\cite{Richard2008}
of the classical ``Discrete Element Method'' method
where Newton's equations of motion for a system of $N$ ``soft'' grains
are integrated. 
This requires giving an explicit expression for the forces that act between grains. 
Such a technique is able to reproduce successfully experimental results for gravity driven flows~\cite{Silbert2001,Taberlet2008,
Taberlet2004b,Tripathi2010,AnkiReddy2010}, sheared systems~\cite{Rycroft2009}, granular materials close to jamming~\cite{Majmudar2007}, silos~\cite{Hirshfeld2001} or rotating drums~\cite{Rapaport2007,Taberlet2004a,Taberlet2006a,Hill2008}.  
The Discrete Element Method  is well known and can be found in many papers~\cite{Silbert2001,Taberlet2008,
Taberlet2004b,Rycroft2009,Majmudar2007,Hirshfeld2001,Rapaport2007,Taberlet2004a,Hill2008,Taberlet2006a,Taberlet2006b,Tripathi2010,AnkiReddy2010}. Therefore,  
we just present here the forces used in this work
(and also in \cite{Silbert_PRL_2005}). 
For the normal force between two overlapping spheres 
we use 
a standard linear spring-dashpot
interaction model~\cite{Luding2008}:
$\mathbf{f}^n=k^n\boldsymbol{\delta^n}-\gamma^n\mathbf{v}^n,$
where $\boldsymbol{\delta^n}$ is the normal overlap, $k^n$ is the spring constant, $\gamma^n$ the damping coefficient and $\mathbf{v}^n$ the normal relative velocity.
The damping models the dissipation characteristic of granular materials. 
Likewise we model the tangential force as a linear elastic and linear dissipative force in the tangential direction:
$\mathbf{f}^t = -k^t\boldsymbol{\delta^t}-\gamma^t\mathbf{v}^t,$
where $k^t$ is the tangential spring constant, $\boldsymbol{\delta^t}$ the tangential overlap, $\gamma^t$ the tangential damping and $\mathbf{v}^t$ the tangential velocity at the contact point.
The magnitude of $\boldsymbol{\delta^t}$ is truncated as necessary to satisfy Coulomb law:  $\left|\mathbf{f}^t\right|\leq\mu\left|\mathbf{{f}}^n \right|$, where $\mu$ is the grain-grain friction coefficient.   %

\textit{Numerical set-up.}
As in \cite{Silbert_PRL_2005},
we simulate
gravity-driven chute flow.
In every way the parameters are the same as in~\cite{Silbert_PRL_2005} except we 
vary the normal stiffness between grains and the height of the flow.
%
The grains are monosized (diameter $d$, mass $m$).
{Unless otherwise specified the number of grains is $N=8000$}.
The chute consists in a $3D$ cell 
whose base is flat and rectangular with size $20d \times 10d$. 
It can be inclined relative to the horizontal by an angle $\theta$ (angle between the horizontal and the long axis of the base) and is periodic in the directions tangent to the base. The bottom of the cell 
is obtained by pouring under gravity $\mathbf{g}$ a large number of grains in the cell ($\theta=0^\circ$) and by fixing those that are in contact with the base.
This disordered layer of fixed grains is sufficient to prevent crystallization throughout the system.  Before any measurements are taken, the inclination is increased to 
$\theta\approx 30^\circ$, causing a rapid flow that erases any influence of the initial state.  The angle $\theta$ is then set to
 a final value that is cited in the captions or text below.
The following
values of the parameters are used:  
$2\times 10^4 \leq k^nd/mg \leq 2\times 10^6$, 
$k^t=2k^n/7$, $\gamma^t=0$ and  $\mu = 0.5$. The value of $\gamma^n$ is adjusted to obtain a normal restitution coefficient $e_n = 0.88$~\cite{Silbert2001}.
We use dimensionless quantities by measuring
distances, times, and elastic constants
respectively, in units of $d$,  $\sqrt{d/g}$ and $mg/d$. 

\begin{figure}[tb]
\begin{center}
\includegraphics*[width=1.0\columnwidth]{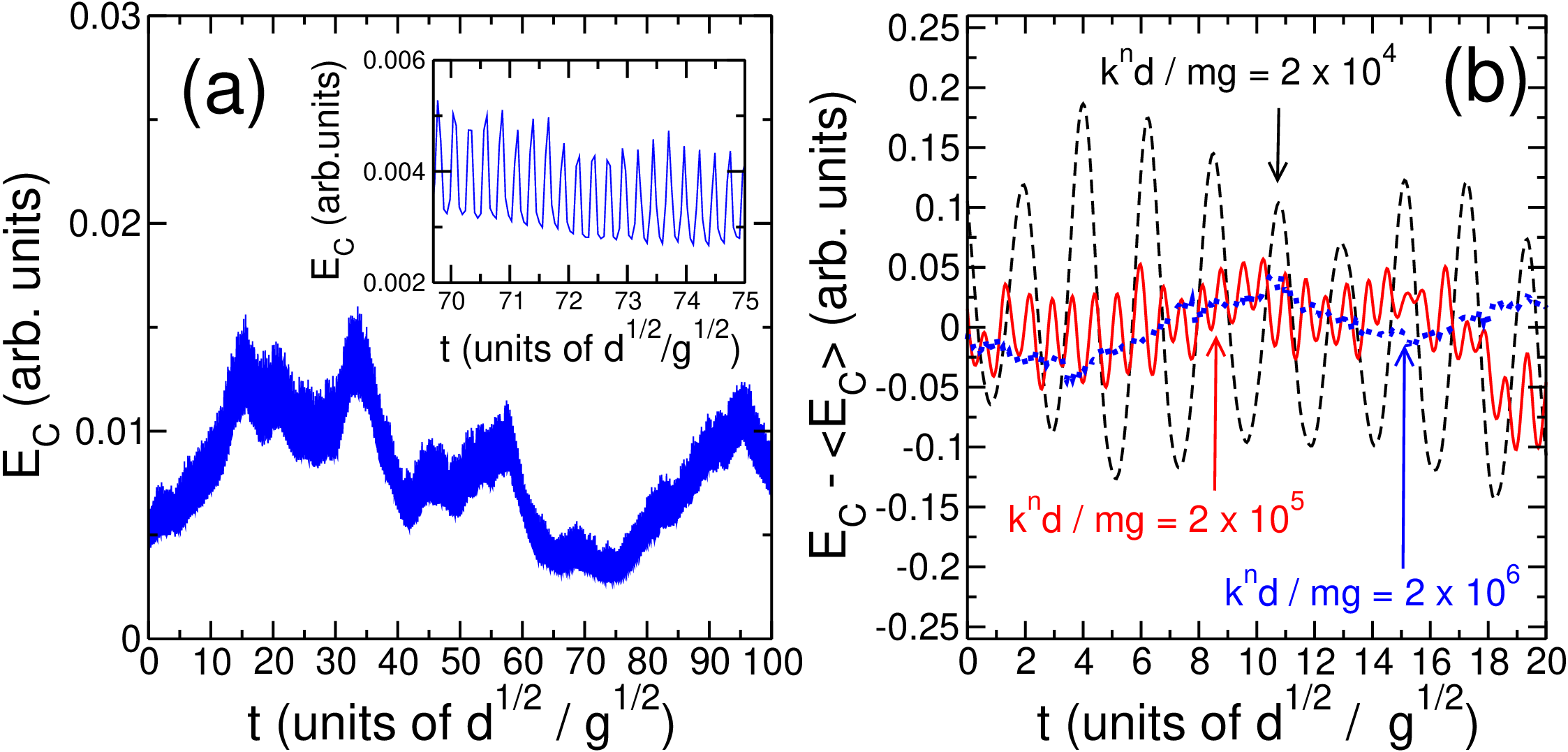}
\caption{\label{fig:EC} (color online)(a) Average kinetic energy per particle for an angle of inclination $\theta = 19.5^\circ$ and $k^nd/mg=2\times10^6$. Fluctuations (main panel) at large time scales and oscillations at small time scales can be observed (inset). (b) Average kinetic energy per particle for 
$\theta = 20^\circ$ (the angle of repose is $19.1^\circ$) and for different values of 
$k_n$.
}
\end{center}
\end{figure}

\textit{Numerical Results.}
In accord with previous results \cite{Silbert_PRL_2005},
we observe an oscillation near the angle of repose
which is defined, for a given flow height, by the angle below which the flow stops.
This motion can be identified through measurements of the total kinetic energy.
Note that in all our simulations the rotational kinetic energy is much lower than the translational one.
As reported in Fig.~\ref{fig:EC}a, the kinetic energy displays regular oscillations (characteristic frequency $\approx \sqrt{g/d}$)  and irregular fluctuations with lower frequency
($\approx 0.1~\mbox{ to }0.2 \sqrt{g/d}$).
%
We next turn our attention to the spring stiffness $k^n$ 
 by studying its effect on
the amplitude and on the frequency of the oscillations.
As reported in Fig.~\ref{fig:EC}b, the oscillations indeed depend on the spring stiffness. 
As $k^n$ increases,
their frequency increases while their amplitude decreases.
\begin{figure}[tb]
\begin{center}
\includegraphics*[width=1.0\columnwidth]{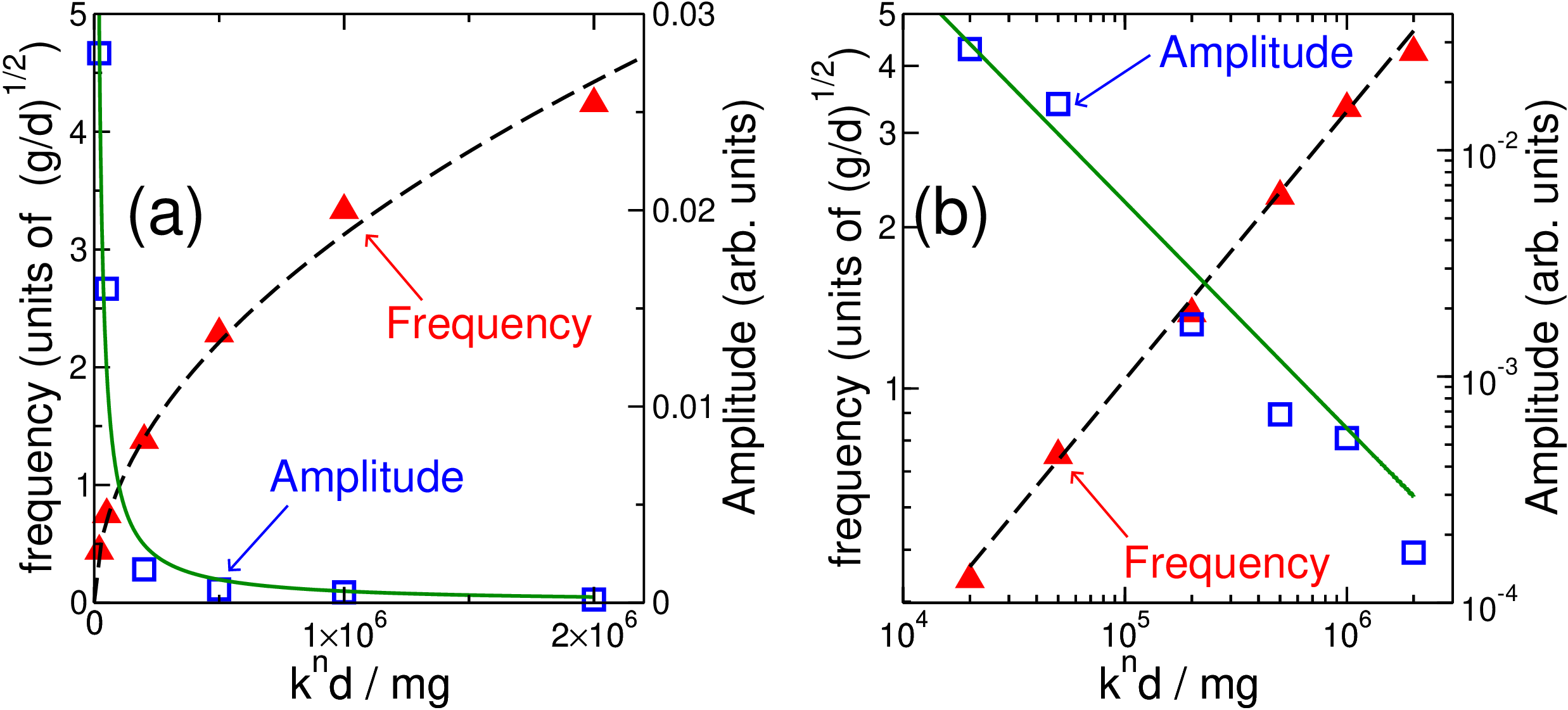}
\caption{\label{fig:freq_vs_kn} (color online) (a) Frequency (cycles per time unit) of the kinetic energy oscillations ($\circ$), corresponding $1/2$ power fit (dashed line), amplitude of the corresponding Fourier component ($\Box$) and corresponding $(k^n)^{-1}$ fit (full line). The angle of inclination is $\theta=20^\circ$. (b) same curves in a log-log plot. }
\end{center}
\end{figure}
Fig.~\ref{fig:freq_vs_kn} shows
the frequency and amplitude of these oscillations
versus spring stiffness.
Frequency scales with $\sqrt{k^n}$,
whereas amplitude scales with $1/k^n$.
In the limit $k^n\to\infty$,
the oscillations disappear.
However, the avalanche
does not disappear in this limit, indicating that the avalanche
and the oscillations are two separate processes.  For this reason,
we do not think that the oscillations should be interpreted as
arising from the jamming transition, or as avalanche precursors.
Now let us consider the oscillation frequency.
It scales as $1/t_c$ 
where $t_c$ is the two-body collision time.
Assuming that a two particle collision is a demi-cycle
of a damped harmonic oscillator leads to
$t_c=\pi/[{2k^n}/{m}- {(\gamma^n})^2/m^2]^{1/2} \approx \pi/(2k^n/m)^{1/2} \propto
1/\sqrt{k^n}$.
This scaling of the frequency suggests
that the oscillations are harmonic vibrations
of a normal mode.
Let us estimate the frequency of the vertical normal modes.
In the simulations,
there are a certain number $n_L$ of
layers of grains resting on the bottom of the chute.
We model the granular bed as a one dimensional chain
of $n_L$ masses connected by linear springs.
This model is the linear harmonic chain,
used in solid state physics as a very elementary model of 
phonons \cite{AshcroftMermin}.
The damping added to the particle interactions
affects short wavelength vibrations most strongly,
but long wavelength ones only weakly \cite{FahrangBook}.
Thus the motion is dominated by the longest possible wavelength.
If the bottom of the chute is considered to be fixed,
and the top surface is free,
the longest wavelength is four times the depth of the layer.
This leads to a frequency
$f_{n_L}/\sqrt{g/d} \approx (1/4n_L)  \sqrt{k^nd/mg}$.
With $n_L=40$ and $k^nd/mg=2\times10^6$,
one obtains $f_{40} \approx 9 \sqrt{g/d}$.
This value is twice as large as the one shown in Fig.~\ref{fig:freq_vs_kn}.
The difference between the prediction and the simulation
is probably caused by the proximity of the stability threshold.
The macroscopic stiffness of a granular packing decreases as
a yield condition is approached.
Here, the yield condition (i.e the avalanche) is reached when $\theta$
equals the angle of repose.
To support this idea, simulations were carried out using exactly the same protocol but with a final angle of $\theta=0^\circ$.  The system comes rapidly to rest,
with the longest-lived motions being   
persistent oscillations at frequencies closer to the predicted values.
\begin{figure}[tb]
\begin{center}
\includegraphics*[width=0.75\columnwidth]{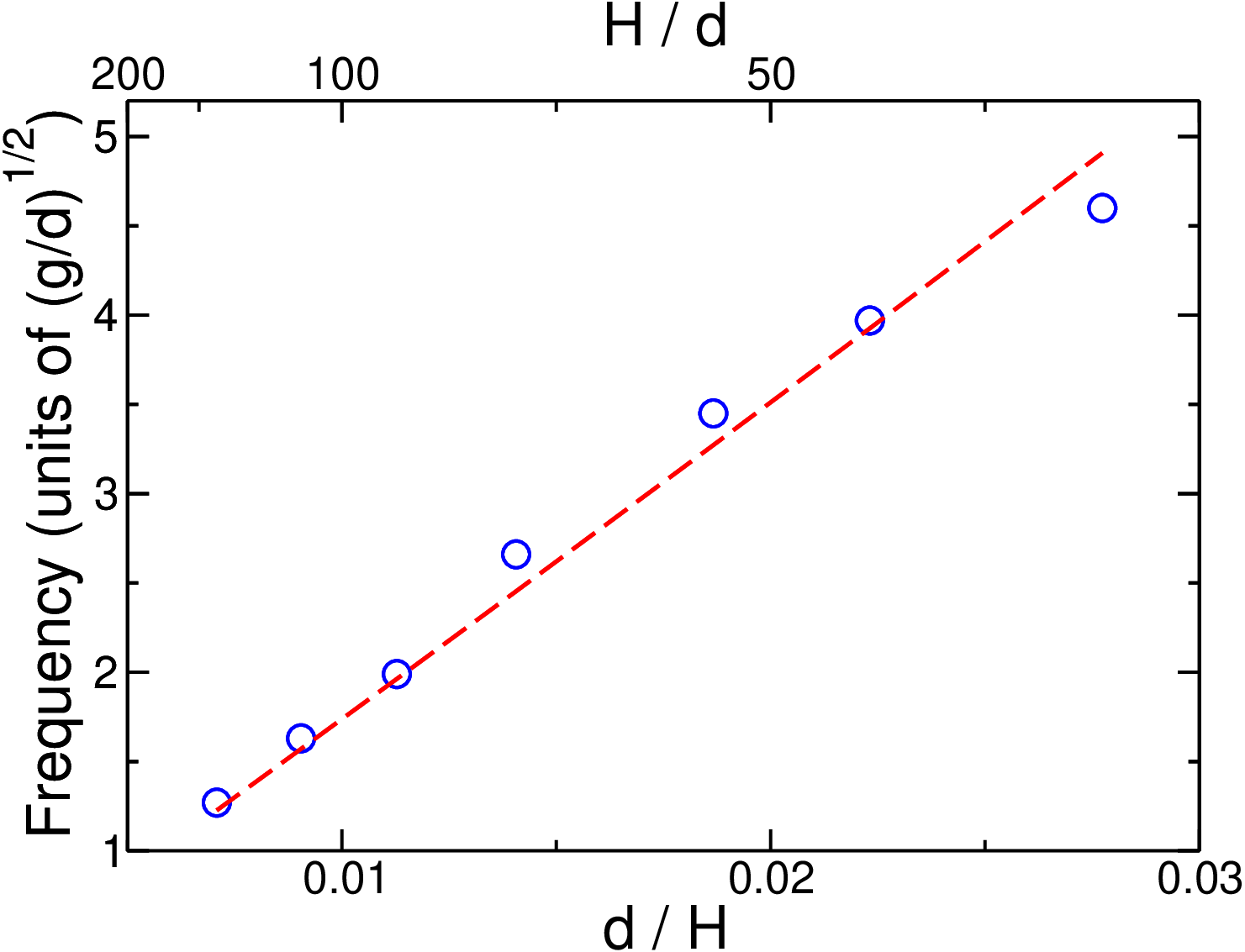}
\caption{\label{fig:freq_vs_N} (color online) Frequency (cycles per time unit) of the kinetic energy oscillations ($\circ$) versus the height of the flow 
and corresponding fit (dashed line).
The angle of inclination is $\theta=20^\circ$ and the normalized stiffness $k^nd/mg=2\times 10^6$.}
\end{center}
\end{figure}
But the central point is that the model
reproduces the correct scaling with stiffness $f_{n_L} \sim \sqrt{k^n}$.
Our results also predict that frequency should
diminish as the granular layer is made deeper: $f_{n_L} \propto 1/n_L$.
To check our analysis we carried out numerical simulations with $k^n d /mg=2\times 10^6$, $\theta=20^\circ$ and for several number of grains {$(8000 < N < 32000)$}.
The results shown in Fig.~\ref{fig:freq_vs_N} confirm
the predicted scaling.
We conclude, therefore, that the oscillations observed in the simulations
are simply harmonic oscillations of the lowest normal mode.

\begin{figure}[tb]
\includegraphics*[width=0.8\columnwidth]{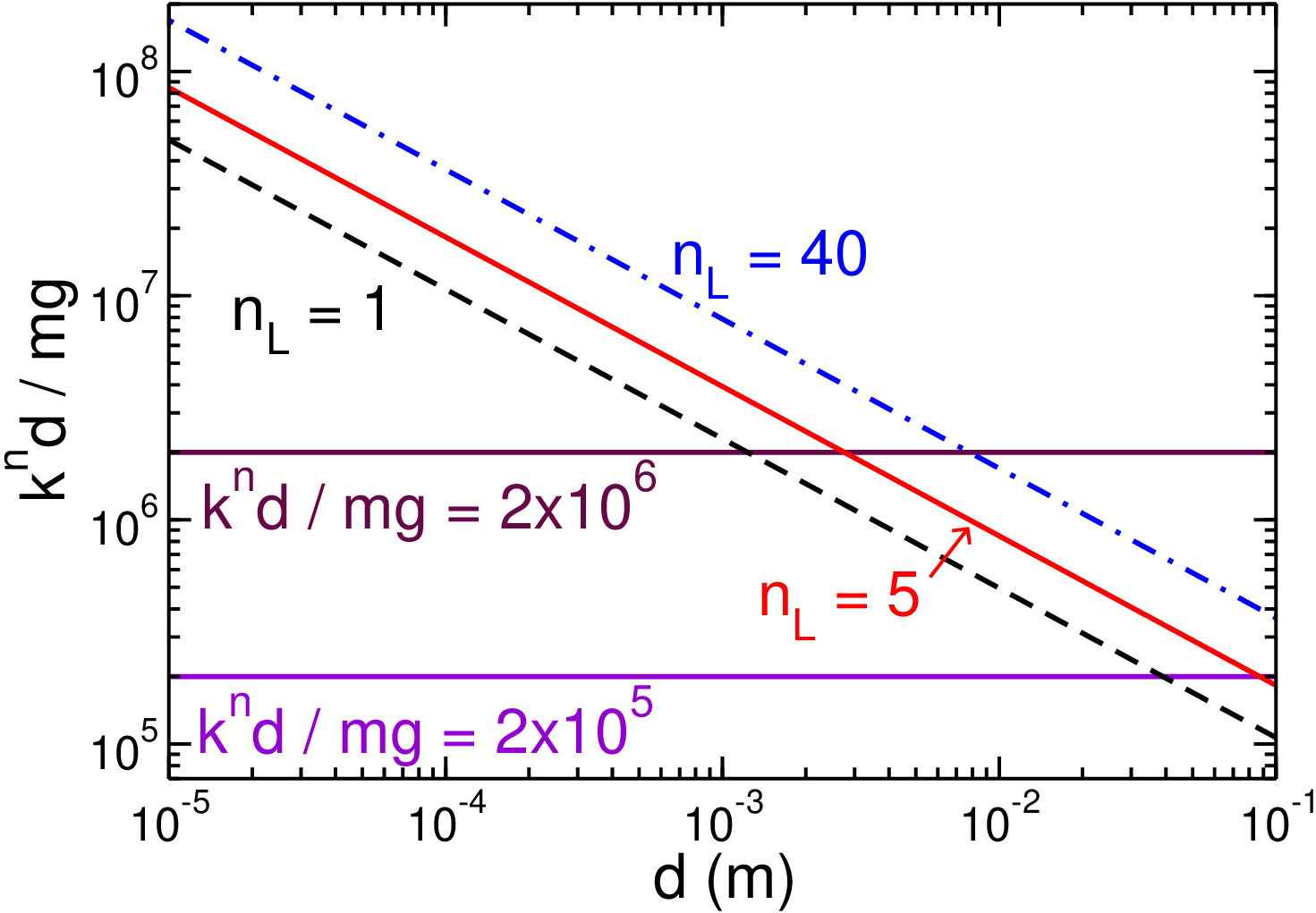}
\caption{(color online) The dimensionless stiffness of the contact as estimated 
by Eqs.~(\ref{eq:stiffness-result}) and (\ref{eq:stiffness-K})
for a variety of diameters $d$ (in meters)
and number $n_L$ of layers of particles.}
\label{fig:stiffness}
\end{figure}

\textsl{Relevance of the numerical results.}
Up to this point,
we have discussed our results solely within the framework of the very idealized model
where a relatively small number of perfect spheres flow down a fixed inclined plane,
interacting via linear spring forces. 
The linear force law allowed us to reproduce the oscillations
under exactly the same conditions as Silbert~\cite{Silbert_PRL_2005},
and also facilitated the analysis of the observed frequencies.
The idealized context of our work is emphasized
by the exclusive use of dimensionless units.
We did not try to give physical values to any parameter such as the particle diameter $d$.
For the remainder of the paper,
we will investigate the suggestion of Mills et al.~\cite{Mills2009}
and discuss the relevance of our findings to ``booming sand''.
This means that we must assign physical values to all quantities.
The most troublesome (and the most important)
parameter is the spring stiffness $k^n$.  Real sand grains have non-linear force-displacement
laws that cannot be characterized by a single spring constant.
We will, nevertheless, try to straddle the difference
between the model and physical system by choosing a single value of $k^n$
relevant to the oscillations.
This approach may be problematic:
Modifying $k^n$ affects the rheology of the flow~\cite{AnkiReddy2010},
even for the highest values of $k^n$ used in this  work~\cite{Tripathi2010}.
To determine the appropriate value of $k^n$,
we will consider a slightly less idealized model
where the grains are spheres made out of an isotropic elastic material
with Young modulus $E$ and Poisson ratio $\nu$.
The contact force between two such spheres is given by the Hertz law:
\begin{equation}
F^n = \frac{E \sqrt{2d}}{3(1-\nu^2)} \delta^{3/2}.
\label{eq:Hertz}
\end{equation}
We will take $k^n$ to be the stiffness seen by the grains
when they oscillate at low amplitude about their equilibrium positions.
We have thus $k^n = dF^n/d\delta = (3/2)F_n/\delta$.
Rearranging this equation, we obtain 
\begin{equation}
F^n = 2 k^n \delta / 3.
\label{eq:stiffness}
\end{equation}
In this equation,
$k^n$ is not a constant: $k^n \propto \delta^{1/2}$.
We seek a typical value of $k^n$ that will be determined by
the typical contact force.
We suppose that this force is of the order of the weight of
a column of $n_L$ grains:
\begin{equation}
F^n = n_Lmg = \pi n_L \rho gd^3/6,
\label{eq:weight}
\end{equation}
where $\rho$ is the density of the material making up the grains.
Eqs.~(\ref{eq:Hertz}), (\ref{eq:stiffness}), and (\ref{eq:weight})
involve the unknowns $F^n$, $k^n$, and $\delta$.
Combining them yields to
\begin{equation}
\frac{k^n d}{mg} = \frac32 n_L \frac{d}{\delta}
= \frac32 \left[ \frac{2 \sqrt{2n_L} E}{\pi(1-\nu^2) \rho dg} \right]^{2/3}.
\label{eq:stiffness-result}
\end{equation}
We will consider modifying $n_L$ and $d$,
while sweeping all the other constants
into a single parameter that depends on the material:
\begin{equation}
\frac{k^n d}{mg} = K \frac{n_L^{1/3}}{d^{2/3}}, \quad
K = \frac32 \left[ \frac{2\sqrt{2}E}{\pi(1-\nu^2)\rho g} \right]^{2/3}.
\label{eq:stiffness-K}
\end{equation}
Putting in values of appropriate for glass:
$g = 10 \mathrm{m}/\mathrm{s}^2$,
$\rho=2.4\times10^3\,\mathrm{kg}/\mathrm{m}^3$,
$\nu=0.2$,
and
$E = 50\,\mathrm{GPa}$,
one obtains
$K = 2.3 \times 10^4 \,\mathrm{m}^{2/3}$.
Various values of the dimensionless stiffness $k^nd/mg$,
are shown in Fig.~\ref{fig:stiffness}.
Note that at $d \approx 1\,\mathrm{mm}$,
even the softest contact (those that support the weight of only one grain)
have a stiffness $k^nd/mg \approx 10^6$,
i.e., equal to the half of the highest value in Fig.~\ref{fig:EC}.
Diameters typical of ``booming dunes'' --
around $200\,\mu \mathrm{m}$ -- lead to even stiffer contacts.
Now let us see whether the oscillation frequencies in the simulation correspond to those of ``singing sand''.
The unit of frequency used here is $\sqrt{g/d}
\approx 220\,\mathrm{Hz}$ for $d=200\,\mu\mathrm{m}$ (which is, as mentioned before, the typical diameter for grains of booming dunes). The observed frequencies in the field and in laboratory are around $90\,\rm{Hz}$, i.e. $0.4\sqrt{g/d}$
(see Fig.~3 of~\cite{Vriend2007} or table I of~\cite{Douady2006}).
According to 
laboratory experiments~\cite{Douady2006}, flow heights are of order of several centimeters.
For $d=200\,\mu\rm{m}$, this corresponds to $H/d \approx 500$.
Extrapolating the data in Fig~\ref{fig:freq_vs_N} to $H/d=500$
leads to frequencies tantalizingly close to those observed.
But Fig.~\ref{fig:freq_vs_N} was obtained for a fixed stiffness
$k^nd/mg=2\times10^6$, much lower than the value 
$k^nd/mg=5.3\times10^7$ predicted by
Eq.~(\ref{eq:stiffness-K}) for $n_L=500$ and $d=200\,\mu\rm{m}$.
If the oscillations are indeed the origin of booming sand,
the effective stiffness of the grains must be much lower than
predicted by Eq.~(\ref{eq:stiffness-K}).
To explain this disagreement, let us determine 
the overlap $\delta$ between two contacting grains
from Hertz theory which assumes those grains as perfect spheres, without any 
surface asperities.
Eqs.~(\ref{eq:Hertz}) and~(\ref{eq:weight}) yield to
\begin{equation}
\delta = \frac{3d^{5/3}n_L^{2/3}}{2K}.
\end{equation}
For $d=200\,\mu\rm{m}$ and $n_L=500$,
we obtain $\delta\approx3\,\mathrm{nm}$.
This is a maximum value of $\delta$,
concerning contacts that support the weight of $500$ grains.
Contacts near 
the free surface of the flow
($n_L<5$) have $\delta<2\,\mathrm{\AA}$.
For such overlaps, 
the contact between two real grains
will be dominated by surface properties that might be quite
different from the bulk properties considered in the Hertz model.
In particular asperities or
a layer of silica gel (as proposed in ~\cite{Goldsack1997})
could significantly reduce the stiffness seen by acoustical waves.
These speculations, however, can only be confirmed by examining
the sand grains themselves.
%
These surface effects could also
account for the rapid increase in sound speed with depth
\cite{Vriend2007}.
The contact stiffness near the dune surface would be 
anomalously soft and fixed by grain surface properties,
whereas deeper in the dune, at higher contact forces,
the stiffness would be much higher and dominated by the bulk properties.


\textsl{Conclusion.}
We have examined the oscillations observed \cite{Silbert_PRL_2005}
in numerical simulations of granular beds near the angle
of repose.
These oscillations are harmonic vibrations
of the lowest vertical mode of the bed,
and their frequency obeys the expected dependency on
particle stiffness and bed depth.
They are not part of the avalanche motion
but may be connected to ``booming dunes'',
if the effective contact stiffness is about $20$
times smaller than expected from the Hertz contact law.
Such a reduction of stiffness is possible
because the Hertz law predicts extremely small grain overlaps,
indicating that the stiffness is dominated by surface
properties instead of bulk ones.
If the oscillations are indeed related to ``booming sand'',
it would mean that the sound originates from a resonance
inside a flowing layer,
similar to the explanation presented in \cite{Douady2006}.
However, several issues remain open.
For example, what excites the oscillations?
One clue is given by the effect of polydispersity:
we performed simulations with 
uniform size distribution 
of width $2d\sigma$, with $\sigma=0.05,\ 0.1$ or $0.2$, and observe that the oscillations disappear for $\sigma=0.2$,
consistent with the observation~\cite{Lindsay1976} that booming sand has a narrow
grain size distribution.
This suggests that the resonance is excited by quasi-periodic
collisions in the shearing layer, consistent with most explanations
that have been presented.

\textsl{Acknowledgments.} We thank J.T. Jenkins, D. Tan for stimulating discussions. This work
is supported by the ANR project STABINGRAM No.
2010-BLAN-0927-01 and by the r\'egion Bretagne (CREATE SAMPLEO). M. T. is supported by the r\'egion Bretagne (ARED grant).

\end{document}